\documentstyle[aps,twocolumn]{revtex}

\def\Rey{{\text\it Re}}          
\def\hatx{\hat{\bf x}}           
\def\lesssim{\,
  \mbox{\raisebox{-0.7ex}{\mbox{$\stackrel{\textstyle<}{\sim}$}}}\;}
\def\etal{\mbox{\it et al.}}
\def\argxt{\!\left({\bf x}, t\right)}
\def\argx{\!\left({\bf x}\right)}
\def\argzeta{\!\left(\zeta\right)}
\def\argphi{\!\left\{\phi\right\}}
\def\argRe{\!\left(\Rey\,\right)}
\def\ceps{c_{\varepsilon}}
\def\cbar{\overline{c_{\varepsilon}}}
\def\Rti{\Rey\to\infty}
\def\clow{\underline{c_{\varepsilon}}}
\def\RES{\Rey_{ES}}
\def\RBi{\Rey_{B}}
\preprint{}
\draft
\title{Variational bound on energy dissipation in turbulent shear flow}
\author{Rolf Nicodemus, Siegfried Grossmann, and Martin Holthaus}
\address{Fachbereich Physik der Philipps-Universit\"at, \\
	Renthof 6, D--35032 Marburg, Germany}
\date{July 25, 1997; revised September 26}
\narrowtext
\begin{document}
\maketitle
\begin{abstract}
We present numerical solutions to the extended Doering--Constantin
variational principle for upper bounds on the energy dissipation rate
in plane Couette flow, bridging the entire range from low to
asymptotically high Reynolds numbers. Our variational bound exhibits
structure, namely a pronounced minimum at intermediate Reynolds numbers,
and recovers the Busse bound in the asymptotic regime. The most notable
feature is a bifurcation of the minimizing wavenumbers, giving rise to
simple scaling of the optimized variational parameters, and of the
upper bound, with the Reynolds number.
\end{abstract}

Rigorous results are rare in the theory of turbulence. There are no
analytical solutions to the Navier--Stokes equations for fully developed
turbulence, and numerical simulations can only cope with flows at
comparatively low Reynolds numbers, due to the enormous number of degrees of
freedom involved. The recent formulation by Doering and Constantin of a
novel variational principle for computing rigorous upper bounds on
quantities characterizing turbulent flows~\cite{DC94,CD95,DC96} has
therefore met with considerable interest. When resorting to a variational
principle, one does no longer aim at solving the equations of motion
exactly, but rather uses these equations to derive inequalities that bound
the relevant quantities, such as the rate of energy dissipation or other
transport properties. The hope is to obtain variational inequalities that
are, on the one hand, still technically manageable, but capture essential
features of the full dynamical fluid system on the other.

This spirit has been very much alive already some 25 years ago in the
Howard--Busse theory~\cite{H72,B70,B78}. The renewed excitement about such
methods stems from an obvious question: Does the new principle provide
new physical insight?

In this Letter, we answer this question for energy dissipation in plane
Couette flow. We have devised a numerical scheme that allows us to exhaust
the Doering--Constantin variational principle, with the extension introduced
in Ref.~\cite{NGH97a}, for the entire range from low to asymptotically high
Reynolds numbers. Intriguingly, we find that the solution to this intricate
problem is organized by a simple feature: the minimizing wavenumbers
bifurcate at a certain intermediate Reynolds number, there\-by determining
the asymptotic scaling of the optimal upper bound on the energy dissipation
rate. As a consequence of this bifurcation, this upper bound exhibits some
structure at about those Reynolds numbers where typical laboratory shear
flows become turbulent.

We start from the equations of motion
\begin{eqnarray}
  \partial_t{\bf u}+{\bf u}{\bf\,\cdot\,\nabla}{\bf u}+{\bf\nabla}p & = &
    \nu\,\Delta{\bf u}, \label{NSE} \\
  {\bf\nabla\,\cdot\,}{\bf u} & = & 0
\end{eqnarray}
for an incompressible fluid confined between two infinitely extended, rigid
plates; $\nu$ is the kinematic viscosity and $p$ the kinematic pressure. The
lower plate, coinciding with the plane $z=0$ of a Cartesian coordinate
system, is at rest, whereas the upper one at $z=h$ is sheared with constant
velocity $U$ in positive $x$-direction. This yields the no-slip boundary
conditions
\begin{equation}
  {\bf u}\!\left(x,y,0,t\right) = {\bf 0}, \quad
    {\bf u}\!\left(x,y,h,t\right) = U\hatx \label{NSB}
\end{equation}
($\hatx$ is the unit vector in $x$-direction); periodic boundary conditions
are imposed in $x$- and $y$-direction. The time-averaged rate of energy per
mass dissipated in the periodicity volume $\Omega$ is given by
\begin{equation}
  \varepsilon_T\equiv\frac{1}{T}\int\limits_{\!0}^{\;T}\!{\rm d}t
  \left\{\frac{\nu}{\Omega}\int\limits_{\Omega}{\rm d}^3\!x
  \left[\sum_{i,j=x,y,z}\!\!\left(\partial_j u_i\right)^2\right]\right\}.
  \label{EPS}
\end{equation}
Our aim is to compute a rigorous upper bound on the long-time limit of the
non-dimensionalized dissipation rate
\begin{equation}
  \ceps\argRe\equiv\lim_{T\rightarrow\infty}\frac{\varepsilon_T}{U^3 h^{-1}},
\end{equation}
where $\Rey=Uh/\nu$ is the Reynolds number.

Let us recall~\cite{J76,DR81} that below the energy stability limit
$\RES\approx 82.65$ we have the exact identity $\ceps\argRe=\Rey^{-1}$.
Moreover, the laminar flow gives rise to the lower bound
$\clow\argRe=\Rey^{-1}$ for all $\Rey$~\cite{DC94}. In order to calculate an
upper bound on $\ceps$ for $\Rey\geq\RES$, we employ the background flow
approach pioneered by Doering and Constantin. Decomposing the actual
velocity field ${\bf u}$ into a stationary, divergence-free ``background
flow'' ${\bf U}$ (not to be confused with the physical mean flow), which has
to carry the physical boundary conditions, and the deviations ${\bf v}$ from
that flow,
\begin{equation}
  {\bf u}\argxt = {\bf U}\argx + {\bf v}\argxt,
\end{equation}
inserting this decomposition into Eq.~(\ref{EPS}), and utilizing the
equations of motion (\ref{NSE})--(\ref{NSB}) when eliminating the deviations
${\bf v}$, one derives an inequality that bounds $\ceps$ in terms of
${\bf U}$~\cite{DC94}. We confine ourselves to background flows of the form
\begin{equation}
  {\bf U}\argx\equiv U\phi\argzeta\hatx,
\end{equation}
where $\zeta\equiv z/h$, and introduce a dimensionless balance parameter
$a>1$, which weights the contribution to $\varepsilon_T$ that stems
from the cross-term containing both background flow and deviations. This
parameter constitutes an additional degree of freedom and thus improves the
bound. The bounding inequality then translates into a variational principle
for $a$ and the profiles $\phi\argzeta$~\cite{NGH97a}:
\begin{equation}
  \ceps\argRe\leq\min_{\phi,\,a>1}
    \left\{\left[1+\frac{a^2}{4\left(a-1\right)}\,
    D\argphi\right]\Rey^{-1}\right\},
  \label{VAP}
\end{equation}
where $D\argphi$ denotes the profile functional
\begin{equation}
  D\argphi\equiv\int\limits_{\!0}^{\;1}\!{\rm d}\zeta
  \left[\phi^{\prime}\argzeta\right]^2-1.
\end{equation}
The essential point now is that the candidate profiles $\phi$ are restricted
by a spectral constraint: for each admissible $\phi$, all eigenvalues
$\lambda$ of the linear eigenvalue problem
\begin{eqnarray}
  \lambda{\bf V} & = & - 2\,h^2\Delta{\bf V} + R\,\phi^{\prime}
    \left(\begin{array}{ccc} 0 & 0 & 1 \\ 0 & 0 & 0 \\ 1 & 0 & 0
      \end{array}\right)
    {\bf V} + {\bf\nabla} P, \nonumber \\
  0 & = & {\bf\nabla\,\cdot\,}{\bf V},
  \label{SPC}
\end{eqnarray}
have to be positive; the eigenvectors ${\bf V}\argx$ have to satisfy
the same homogeneous boundary conditions as the deviations ${\bf v}\argxt$.
The number $R$ introduced here is the rescaled Reynolds number,
\begin{equation}
  R\equiv\frac{a}{a-1}\,\Rey. \label{RSC}
\end{equation}

Computationally, the main task is to calculate, for every single candidate
profile $\phi$, that rescaled Reynolds number $R_c\argphi$ for which the
lowest eigenvalue of (\ref{SPC}) passes through zero. When $a$ is adjusted
such that the r.h.s.\ of the inequality (\ref{VAP}) is minimized for fixed
$\phi$ and $\Rey$, subject to the condition~(\ref{RSC}), this $\phi$ produces
via (\ref{VAP}) an upper bound on $\ceps$ for all $\Rey\leq R_c\argphi$.
That is, $R_c\argphi$ marks the highest $\Rey$ up to which $\phi$ is
admissible to the variational principle (\ref{VAP}). Exploiting the periodic
boundary conditions, each eigenvalue $\lambda$ is labeled by two
wavenumbers, $k_x$ and $k_y$. We first keep these wavenumbers fixed and
compute that number $R=R_0\argphi\!\left(k_x,k_y\right)$ for which the
lowest eigenvalue $\lambda_{k_x,k_y}$ becomes zero, and then determine
$R_c\argphi$ by minimizing over all $k_x$ and $k_y$. The details of this
tedious procedure will be given in Ref.~\cite{NGH97b}.

We use variational profiles of the type displayed in Fig.~\ref{F_1}:
boundary layer segments of width $\delta$, modeled by polynomials of order
$n$, are connected by a linear piece with slope $p$, such that the
profiles are $n-1$ times continuously differentiable at the matching points
$z/h=\delta$ and $z/h=1-\delta$. Thus, we have the three variational
parameters $\delta$, $n$, and $p$. These profiles have not been chosen ad
libitum, but have been shown to correctly capture the asymptotics of the
dissipation bound in a closely related model problem~\cite{NGH97c}.

The optimized profiles resulting from the variational principle are depicted
in Fig.~\ref{F_2}. Considering the response of the optimized profile
parameters to the increase of $\Rey$, we can clearly distinguish five
different regimes. (i) Up to $\Rey=\RES$ we have the laminar regime with
$\phi\argzeta=\zeta$. (ii) For $\RES\leq\Rey\leq\Rey_1\approx 160$ the
laminar profile is deformed. The parameters $\delta$ and $n$ remain at their
values $0.5$ and $3$, respectively, while the slope $p$ decreases from $1$
to almost $0$. (iii) In the following regime,
$\Rey_1\leq\Rey\leq\Rey_2\approx 670$, boundary layers develop. Here $n$
still remains fixed, $p$ increases again, while $\delta$ decreases to its
minimal value $0.14$. (iv) Then, for $\Rey_2\leq\Rey\leq\Rey_3\approx 1845$,
$n$ increases dramatically from $3$ to $34$, thus steepening the profile in
the immediate vicinity of each boundary, thereby effectively generating a
new internal boundary layer within each boundary segment. As a consequence,
$\delta$ increases back to $0.5$, so that the boundary segments finally join
together. (v) For $\Rey\geq\Rey_3$ the variational parameters obey simple
scaling laws: $\delta=0.5$ remains fixed, the profile slope at the boundary
becomes
\begin{equation}
  \phi^{\prime}\!\left(0\right)\sim n\propto\Rey,
  \label{PPB}
\end{equation}
while the slope at the midpoint is given by
\begin{equation}
  \phi^{\prime}\!\left({\textstyle\frac{1}{2}}\right)=p\propto\Rey^{-1}.
  \label{PPM}
\end{equation}
The size of the internal boundary layers, i.e., the range close to the
boundaries where the optimized profiles are steep, is measured by
$1/\phi^{\prime}\!\left(0\right)$ and hence proportional to $\Rey^{-1}$.
It needs to be emphasized, however, that the optimal variational profiles
do not necessarily resemble the physically realized mean flow profiles.

A key for understanding the above scaling is provided by the behavior of the
minimizing wavenumbers. Remarkably, the minimizing $k_x$ is zero for
all $\Rey$, reflecting the fact that there is no distinguished length
scale in shear direction. Even more important, the minimizing $k_y$
{\em bifurcates\/} at
\begin{equation}
  \RBi\approx 460,
\end{equation}
as shown in Fig.~\ref{F_3}. This means that for $\Rey>\RBi$ the two lowest
eigenvalues of Eqs.~(\ref{SPC}) are degenerate and pass through zero
simultaneously; it is a key property of the variational principle that both
corresponding minima of $R_0\argphi\!\left(0,k_y\right)$ adopt identical
values. The lower $k_y$-branch ($k_{y,1}$) approaches a constant value for
$\Rti$, namely, the very same $k_y$ that also determines the energy
stability limit, whereas the upper branch ($k_{y,2}$) scales proportionally
to $\Rey$. Both findings are intimately connected to the power-law behavior
of the profile expressed by Eqs.~(\ref{PPB}) and (\ref{PPM}). For Reynolds
numbers above $\RBi$ two different length scales, $k_{y,1}^{-1}$ and
$k_{y,2}^{-1}$, appear on the stage, which can be identified~\cite{NGH97b}
as the extension of the profile's flat part in the interior and the
effective widths of the (internal) boundary layers. The regime (iii) can
thus be considered as a cross-over regime from deformed profiles with no
definite length scales to profiles with well-developed boundary layers.

The optimal variational bound on the energy dissipation rate $\ceps$ that
results from these ingredients is depicted in Fig.~\ref{F_4}. Immediately
above $\RES$ this upper bound $\cbar$ increases, exhibits a maximum, and
then decreases roughly proportional to $\Rey^{-1/4}$ in regime (iii). The
occurrence of the bifurcation changes this behavior: the bound passes
through a distinct minimum at those $\Rey$ where the internal boundary
layers start to develop, and finally ascends to its asymptotic value
\begin{equation}
  \lim_{\Rti}\cbar\argRe=0.01087(1).
\end{equation}
This asymptotic bound has to be compared to the bound calculated by Busse in
the framework of his Optimum Theory~\cite{B70,B78}, which refers to the
limit $\Rti$ and reads
\begin{equation}
  \lim_{\Rti}\ceps\argRe\lesssim 0.010. \label{CBU}
\end{equation}
The prediction that the bound furnished by the variational principle
(\ref{VAP}) should coincide in the limit $\Rti$ with the Busse bound
(\ref{CBU}) has been made by Kerswell~\cite{K97}, who could bring the
variational principle into a form to which Busse's so-called multi-$\alpha$
solutions can be applied. Thus, Kerswell's conclusion had to rely on all
assumptions inherent in the Optimum Theory, whereas we have actually
constructed a rigorous solution to the variational principle (\ref{VAP}). In
this way, we could not only confirm the correctness of Busse's asymptotic
result, but provide a rigorous bound on $\ceps$ for {\em all\/} $\Rey$. Our
bound shows non-trivial structure: a pronounced minimum followed by a
$\Rey$-range between $1000$ and $1800$ in which the bound's curvature changes
its sign. This occurs in regime (iv), i.e., at about those $\Rey$ where one
observes the onset of turbulence in typical laboratory shear flows.


In order to illustrate the importance of the proper implementation of the
spectral constraint, we compare in Fig.~\ref{F_5} the bound obtained in this
work to the bound derived previously by Doering and
Constantin~\cite{DC94,DC92} with the help of elementary functional
estimates. These estimates are convenient to evaluate, but over-satisfy the
spectral constraint and result in the structureless bound
$\ceps\argRe\leq 1/\left(8\sqrt{2}\right)\approx 0.088$ for
$\Rey\geq 8\sqrt{2}$. The improvements due to both the introduction of the
balance parameter $a$ and the evaluation of the less restrictive actual
spectral constraint amounts to almost an order of magnitude, namely to a
factor of $8$ in the asymptotic regime. All upper bounds in Fig.~\ref{F_5}
become asymptotically independent of the Reynolds number, which means that
they are in accordance with the scaling implied by classical turbulence
theories.

To conclude, we have found a scheme for solving the extended
Doering--Constantin variational principle~\cite{DC94,NGH97a} for
upper bounds on energy dissipation in plane Couette flow. The bound
established here is rigorous for all Reynolds numbers, recovers the Busse
bound in the asymptotic regime, and represents in the important regime of
intermediate $\Rey$ the best upper bound that has been calculated so far.
Moreover, we have pinned down the feature that organizes the solutions to
the variational principle, namely the bifurcation of the minimizing
wave\-numbers illustrated in Fig.~\ref{F_3}. Based on these results, it is now
possible to obtain similar bounds of comparable quality also for other flows
of interest.

However, comparing the bound with the experimental data indicated in
Fig.~\ref{F_5}, it is clear that this is not the end of the story. The
variational bound lies an order of magnitude above the dissipation rates
measured by Rei\-chardt~\cite{R59} and Lathrop \etal\/~\cite{LFS92},
and these data do not seem to approach a constant value, but still decrease
with $\Rey$. If one could improve the variational principle such that this
decrease is captured, and if it did persist asymptotically, one could make
definite statements about possible intermittency corrections to classical
scaling~\cite{G95}. To proceed along these lines is possibly one of the
greatest challenges in the rigorous theory of turbulence.

This work was supported by the Deutsche Forschungsgemeinschaft via the
Sonderforschungsbereich ``Nicht\-lineare Dynamik'', SFB~185, and by the
German-Israeli-Foundation (GIF).

\begin{figure}
  \caption[Fig~1]{Variational test profiles $\phi\argzeta$ versus
    $\zeta=z/h$.}
  \label{F_1}
\end{figure}

\begin{figure}
  \caption[Fig~2]{Metamorphosis of the optimized variational profiles with
    the increase of the Reynolds number. We have depicted the most important
    $\Rey$-range on a logarithmic scale, beginning with the energy stability
    limit $\RES$ and ending in the scaling regime $\Rey>\Rey_3$.}
  \label{F_2}
\end{figure}

\begin{figure}
  \caption[Fig~3]{Minimizing wavenumber(s) $k_y$ belonging to the upper
    bound on $\ceps$ displayed in Fig.~\ref{F_4}. The wavenumbers of the
    upper branch correspond to the inverse boundary layer thickness of the
    profiles, while those of the lower branch reflect the inverse extension
    of the flat interior profile part.}
  \label{F_3}
\end{figure}

\begin{figure}
  \caption[Fig~4]{Bounds on $\ceps$ for the plane Couette flow. Points
    denote the upper bound $\cbar\argRe$ derived from the variational
    principle (\ref{VAP}); the solid line on the left is the lower bound
    $\clow\argRe=\Rey^{-1}$. The asymptotic value of the upper bound,
    $\lim_{\Rti}\cbar\argRe=0.01087(1)$, lies slightly above, but within
    the uncertainty span of Busse's asymptotic result (\ref{CBU}).}
  \label{F_4}
\end{figure}

\begin{figure}
  \caption[Fig~5]{A synopsis of bounds on $\ceps$ and experimental data.
    --- Slanted straight line on the left:
      lower bound $\clow\argRe=\Rey^{-1}$.
    --- Topmost horizontal line:
      upper bound obtained by Doering and Constantin in
      Refs.~\cite{DC94,DC92} with the help of an over-restrictive profile
      constraint and piecewise linear profiles; $\cbar\argRe\approx 0.088$
      for $\Rey>11.32$.
    --- Heavy dots:
      upper bound obtained in this work from the variational
      principle (\ref{VAP}) supplemented by the actual spectral constraint
      (\ref{SPC}), cf.\ Fig.~\ref{F_4}; $\cbar\argRe\rightarrow 0.01087(1)$.
    --- Joining dashed line:
      asymptotic upper bound (\ref{CBU}) derived by Busse in
      Refs.~\cite{B70,B78}; $\cbar\argRe\rightarrow 0.010(1)$. The
      shaded area denotes the estimated uncertainty of this bound.
    --- Triangles:
      experimental dissipation rates for the plane Couette flow
      measured by Rei\-chardt~\cite{R59}.
    --- Circles:
      experimental dissipation rates for the Taylor--Couette system with
      small gap as measured by Lathrop, Fineberg and Swinney~\cite{LFS92}.}
  \label{F_5}
\end{figure}

\end{document}